\documentclass{svproc}

\usepackage[font={footnotesize}]{caption}
\usepackage{amssymb}
\usepackage{amsmath}
\usepackage{xcolor}

\numberwithin{equation}{section}

\usepackage{graphicx}

\usepackage{color}
\usepackage[affil-it]{authblk}
\newcommand{\rref}[1]{(\ref{#1})}

\def\fddd#1#2{\displaystyle{\frac{\delta #1}{\delta #2}}}

\newcommand{\dsl}[1]{{\displaystyle{#1}}}
\renewcommand{\sfdefault}{bch}

\def\sbarr{\hbox{{\fontfamily{\sfdefault}\selectfont {\scriptsize I}\hskip -.25ex {\scriptsize R}}}}

\newcommand{\RR}{{\sbarr}}

\newcommand{\MI}{{{\mathcal{I}}}}
\newcommand{\MM}{{{\mathcal{M}}}}

\usepackage{graphicx}

\newcommand{\ou}{{\overline{u}}}

\newcommand{\oet}{{\overline{\zeta}}}

\newcommand{\D}{\mathrm{d}}
\newcommand{\drho}{\rho_{{}_\Delta}}

\usepackage{url}

\begin{document}
\mainmatter              
\title{A Hamiltonian set-up for  4-layer \\ density stratified Euler fluids}
\author{R.\ Camassa\inst{1} \and G.\ Falqui\inst{2,4,5} \and
G.\ Ortenzi\inst{2,5,6} \and M.\ Pedroni\inst{3,5} \and T.T.\ Vu Ho\inst{2,5}}
\authorrunning{R. Camassa et al.} 
%
\tocauthor{R.Camassa, G.Falqui, G.Ortenzi, M.Pedroni, T.T Vu Ho}
\institute{University of North Carolina at Chapel Hill, Carolina Center for Interdisciplinary
Applied Mathematics, Department of Mathematics, Chapel Hill, NC 27599, USA\\
\email{camassa@amath.unc.edu}\\ 
\and
Department of Mathematics and Applications, University of Milano-Bicocca,
Via Roberto Cozzi 55, I-20125 Milano, Italy
\\
\email{gregorio.falqui@unimib.it, giovanni.ortenzi@unimib.it, t.vuho@campus.unimib.it}
\and
Dipartimento di Ingegneria Gestionale, dell’Informazione e della Produzione,
Universit\`a di Bergamo, Viale Marconi 5, I-24044 Dalmine (BG), Italy\\
\email{marco.pedroni@unibg.it}
\and
SISSA, via Bonomea 265, I-34136 Trieste, Italy
\and
INFN, Sezione di Milano-Bicocca, Piazza della Scienza 3, I-20126 Milano, Italy
\and 
Dipartimento di Matematica ``Giuseppe Peano'', \\Universit\`{a} di Torino, 10123 Torino, Italy\\
\email{giovanni.ortenzi@unito.it}
}
\maketitle              

\begin{abstract}

By means of the Hamiltonian approach to two-dimensional wave motions in heterogeneous fluids proposed by 
Benjamin \cite{Benjamin} we derive a natural Hamiltonian structure for ideal fluids, density stratified in four homogenous layers, constrained in a channel of fixed total height and infinite lateral length. We derive the Hamiltonian and the equations of motion  in the dispersionless long-wave limit, restricting ourselves to the so-called Boussinesq approximation. The existence  of special symmetric solutions, which generalise to the four-layer case the ones obtained in~\cite{VirMil19} for the three-layer case, is examined.

%
\end{abstract}
\section{Introduction}
Density stratification in incompressible fluids is an important aspect of fluid dynamics, and plays an important role in 
variety of phenomena occurring in both the ocean and the atmosphere. In particular, displacement of fluid parcels from their neutral
buoyancy position within a density stratified flow can result in internal wave motion.
Effective one-dimensional models (in particular, their quasi-linear limit) were introduced to study these phenomena, and were the subject of  a number of investigations (see, e.g.,  \cite{OVS79,Weakly,Fully,LT,Chum08,Chum09,Du16,Eletal17} and references  therein).
Although most of the theoretical and numerical results that can be found in the literature are focussed on the $2$-layer case, multiply-layered fluid configurations appear as effective models of physical phenomena, e.g., in the atmosphere or in mountain lakes. The extension to the $n>2$ layers case can also be seen as a refined approximation to the  real-world continuous stratification of incompressible fluids.
%

\begin{figure}[t]
\begin{center}
\includegraphics[width=.7 \textwidth]{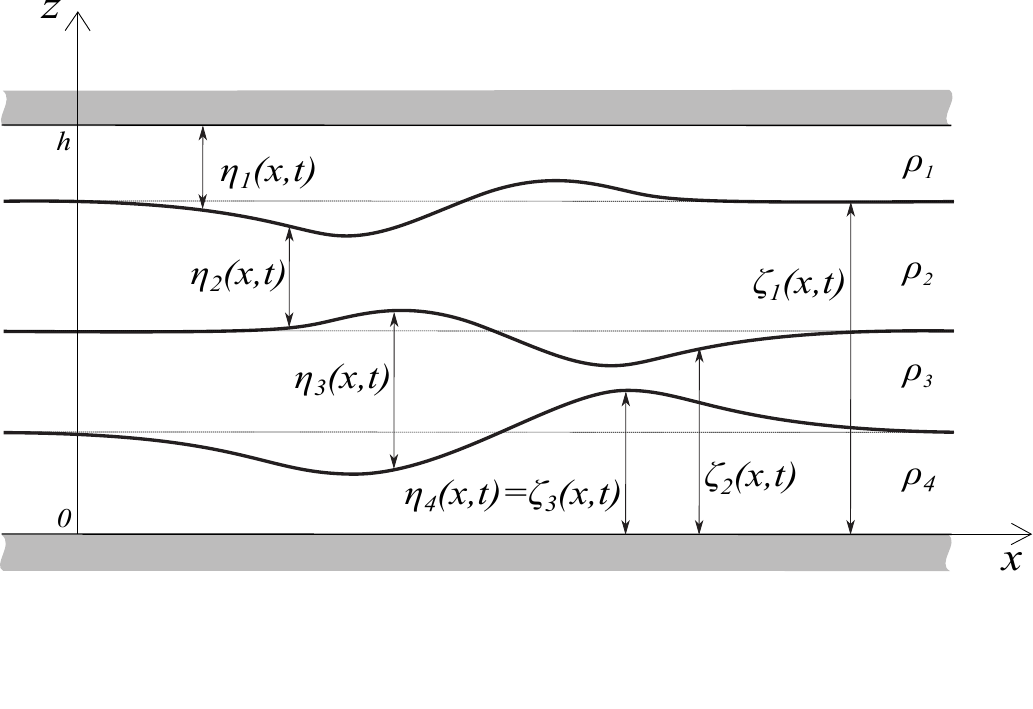}
\caption{Four-layer rigid lid 
setup and relevant notation:  $\zeta_i$ are the surface heights and $\eta_i$ are the layer thicknesses.}
\label{F1}
\end{center}
\end{figure}

The focus of
 the present paper is on the dynamics of an ideal (incompressible, inviscid) stably stratified fluid consisting  of 
$4$~layers of constant density $\rho_1<\rho_2<\rho_3<\rho_4$, confined in a channel of fixed height $h$ (see Figure \ref{F1} for a schematic of our setup), and, in particular, on its Hamiltonian setting.
 This will be obtained by a suitable reduction of the Hamiltonian structure introduced by 
Benjamin \cite{Benjamin} in the study of general density stratifications for Euler fluids in $2$ dimensions.

We shall follow the approach set forth in  our recent paper \cite{Paper}, where the $3$-layer case was considered by extending to the multiple layer case a technique introduced in~\cite{CFO17}.
In particular, after having discussed in details the construction of the Hamiltonian operator for an  effective $1$D model, we shall consider the so-called Boussinesq limit of the system, and explicitly determine its Hamiltonian structure and Hamiltonian functional, as well as point out the existence of special symmetric solutions.

Our mathematical model is based on some simplifying hypotheses. At first, we assume that an inviscid model suffices to capture the essential features of the dynamics since the scales associated with internal waves are large, and consequently the Reynolds number is typically high ($> 10^5$). Although in the ocean and the atmosphere (as well as in laboratory experiments)  the density stratification arises as a consequence of diffusing quantities such as temperature and salinity, we can neglect diffusion and mixing since  the time scales associated with diffusion processes are far larger than the time scale of internal wave propagation. Finally, we use the rigid lid assumption for the  upper surface since  the scales associated with internal wave-motion are greatly exceeding the scales of the surface  waves (see, e.g., \cite{VSH05} for further details on these assumptions).

The Hamiltonian  $4$-layer model herewith discussed  is a natural extension of the $2$ and $3$-layer model. Indeed, when two adjacent densities are equal (and as a consequence the relative interface becomes meaningless) we fully recover the dynamics of the $3$ layer model (see, e.g., \cite{VirMil19,Paper}). Similarly, the $3$-layer model reduces to the ordinary $2$-layer model when two mass densities coincide.

The layout of the paper is the following.
In Section \ref{Sect-1} we briefly review the Hamiltonian representation for $2$-dimensional incompressible Euler fluids of \cite{Benjamin}. Section~\ref{HR} is devoted to a detailed presentation of our Hamiltonian reduction scheme, which endows the dynamics of the set of $4$-layer stratified fluids with a 
natural Hamiltonian structure. In Section~\ref{redham} we compute the reduced Hamiltonian and the ensuing equations of motion, confining ourselves to the case of the so-called Boussinesq approximation. In Section~\ref{SySo} a class of special evolutions, selected by a symmetry of the Hamiltonian, is found and briefly examined.

\section{The 2D Benjamin model for heterogeneous fluids in a channel}\label{Sect-1}
Benjamin \cite{Benjamin} proposed and discussed a  set-up for the Hamiltonian formulation of an incompressible stratified Euler system in $2$ spatial dimensions, 
which we  hereafter summarize for the reader's convenience.
 
 The  Euler equations for a perfect inviscid and incompressible but heterogeneous fluid in 2D, 
subject to gravity $-g\mathbf{k}$, are usually written for the the density $\rho(x,z)$ and the velocity field $\mathbf{u}=(u,w)$  as
\begin{equation}
\label{EEq}
 \begin{split}
 & \frac{D \rho}{D t}=0, \qquad \nabla \cdot \mathbf{u} =0, \qquad \rho \frac{D \mathbf{u} }{D t} + \nabla p + \rho g \mathbf{k}=0 \\
  \end{split}
\end{equation}
together with appropriate boundary conditions, 
where, as usual, $D/Dt=\partial/\partial t+\mathbf{u}\cdot\nabla$ is the material derivative. 

Benjamin's contribution was to consider, as  basic variables for the evolution of such a fluid, the density $\rho$ together with the ``weighted vorticity" $\Sigma$ defined by
\begin{equation}
\label{Sigmadef}
\Sigma =\nabla\times (\rho\mathbf{u})=(\rho w)_{x}-(\rho u)_{z}. 
\end{equation}
The equations of motion for these two fields, ensuing from the Euler equations for incompressible fluids, are 
\begin{equation}
\label{eqsr}
\begin{array}{l}
\rho_t+u\rho_x+w\rho_z =0\\
\noalign{\medskip}
\Sigma_t+u\Sigma_x +w\Sigma_z +\rho_x\big(gz-\frac12(u^2+w^2)\big)_z+\frac12\rho_z\big(u^2+w^2\big)_x=0\, .
\end{array}
\end{equation}
They can be written in the form
\begin{equation}
 \label{heq}
{\rho_{t}}=-\left[\rho,  \dsl{\fddd{H}{\Sigma}}\right] \, , \qquad 
\Sigma_t= -\left[\rho,  \dsl{\fddd{H}{\rho}}\right]-\left[\Sigma, \dsl{\fddd{H}{\Sigma}}\right] \, ,
\end{equation}
where, by definition, the bracket is $[A, B] \equiv A_xB_z-A_zB_x$, and the  functional 
\begin{equation}
\label{ham-ben}
H= \int_\mathcal{D} \rho\left(\frac{1}{2} |\textbf{u} |^2+g z\right)\,{\rm d}x\,{\rm d}z 
\end{equation}
is simply given by the sum of the kinetic and potential energy, $\mathcal{D}$ being the fluid domain.
The most relevant feature of this coordinate choice is that $(\rho,\Sigma)$ are physical, directly measurable, variables.
Their use, though confined to the 2D case with the above definitions, allows one to avoid  the introduction of Clebsch 
variables (and the corresponding subtleties associated with gauge invariance and  limitations of the Clebsch potentials) 
which are often used in the Hamiltonian formulation of both 2D and the general $3D$ case. (see, e.g., \cite{Z85}).\\

As shown by Benjamin, equations (\ref{heq}) are a Hamiltonian system with respect to a 
Lie-theoretic Hamiltonian structure, that is, 
they can be written as
\[
 \rho_t=\{\rho, H\}
,\qquad \Sigma_t=\{\Sigma, H\},
\]
for the Poisson bracket defined by the Hamiltonian operator
\begin{equation}\label{B-pb}
J_B=-
\left(\begin{array}{cc}
       0 & \rho_x \partial_z -\rho_z \partial_x \\ 
       \rho_x \partial_z -\rho_z \partial_x & \Sigma_x \partial_z -\Sigma_z \partial_x
      \end{array}
\right).
\end{equation}

\section{The Hamiltonian reduction process}\label{HR}

As mentioned in the Introduction, we shall consider special stratified fluid configurations, consisting of a fluid with $n=4$~layers
of constant density $\rho_1<\rho_2<\rho_3<\rho_4$ and respective thicknesses $\eta_1$, $\eta_2$, $\eta_3$, $\eta_4$, confined in a channel of fixed height $h$.
We  define, as in Figure \ref{F1}, the locations of the interfaces at $z=\zeta_k$, $k=1,2,3$, related to the thickness $\eta_j$ by 
\begin{equation}
\zeta_3=\eta_4,\quad  \zeta_2=\eta_4+\eta_3, \quad \zeta_1=\eta_4+\eta_3+\eta_2. 
\end{equation}
The velocity components  in each layer are denoted by $\left(u_i(x,z), w_i(x,z)\right)$, $i=1,\dots,4$.
By means of the Heaviside $\theta$  and  Dirac $\delta$ generalized functions, a four-layer fluid configuration can be described within Benjamin's setting as follows.
First, the $2D$ density and velocity variables can be written as 
\begin{equation}\label{rhouw}\begin{split}
&\rho (x,z)=\rho_{4}+(\rho_{3}-\rho_{4})\theta (z-\zeta_{3})+(\rho_{2}-\rho_{3})\theta(z-\zeta_{2})+(\rho_{1}-\rho_{2})\theta(z-\zeta_{1})
\\
&u (x,z)=u_{4}+(u_{3}-u_{4})\theta (z-\zeta_{3})+(u_{2}-u_{3})\theta(z-\zeta_{2})+(u_{1}-u_{2})\theta(z-\zeta_{1})
\\
&w (x,z)=w_{4}+(w_{3}-w_{4})\theta (z-\zeta_{3})+(w_{2}-w_{3})\theta(z-\zeta_{2})+(w_{1}-w_{2})\theta(z-\zeta_{1})\, . \end{split}
\end{equation}
Thus, the density-weighted vorticity $\Sigma =(\rho w)_{x}-(\rho u)_{z}$ can be computed as
\begin{equation}
\begin{split}
    \Sigma=&\sum_{j=1}^3\left(\rho_{j+1}\Omega_{j+1}-  \rho_{j}\Omega_{j}\right) \theta(z-\zeta_j)  +\rho_{4}\Omega_{4}\\
    +&\sum_{j=1}^3\left( (\rho_{j+1}u_{j+1}-\rho_{j}u_{j})+(\rho_{j+1}w_{j+1}-\rho_{j}w_{j}){\zeta_j}_x\right)\delta (z-\zeta_{j})\, , 
    \end{split}
\end{equation}
where $\Omega_{i}={w_i}_x-{u_i}_z$ for $i= 1,\dots,4$ are the  bulk  vorticities.

%
%
%
Next, we assume the bulk motion in each layer to be irrotational, so that $\Omega_{i}=0 $ for all $ i=1,\dots,4$. 
Thus the density weighted vorticity is explicitly given by
\begin{align}
\label{sigma-ham}
    \Sigma=&\left((\rho_{4}u_{4}-\rho_{3}u_{3})+(\rho_{4}w_{4}-\rho_{3}w_{3}){\zeta_3}_x \right) \delta(z-\zeta_{3}) \nonumber \\
    &+\left((\rho_{3}u_{3}-\rho_{2}u_{2})+(\rho_{3}w_{3}-\rho_{2}w_{2}){\zeta_2}_x \right) \delta(z-\zeta_{2}) \\ 
    &+\left((\rho_{2}u_{2}-\rho_{1}u_{1})+(\rho_{2}w_{2}-\rho_{1}w_{1})\zeta_{1x} \right) \delta(z-\zeta_{1})\,.\nonumber
\end{align}
A further assumption we make right from the outset is that of the long-wave asymptotics, with small parameter
$\epsilon=
h/L\ll 1$, $L$ being a typical horizontal scale of the motion such as wawelength. 
This assumption 
implies (see, e.g., \cite{Fully} for further details)
that at the leading order as $\epsilon \to 0$ we have
$$u_{i} \sim {\ou_{i}}\,,\qquad w_{i} \sim 0\,, $$
i.e., we can neglect the vertical velocities $w_{i}$ and trade the horizontal velocities $u_{i}$ with their layer-averaged counterparts,
\begin{equation}
\ou_{i}= \frac1{\eta_i}\int_{\zeta_{i}}^{\zeta_{i-1}} u(x,z)\D z,\, \text{where } \zeta_0\equiv h,\,\,\zeta_4\equiv 0\, .
\end{equation}
Hence, from \rref{sigma-ham} and recalling the first of (\ref{rhouw}), we obtain 
\begin{equation}\label{I-sigma-rho}
\begin{split}
     \rho (x,z)&=\rho (x,z)=\rho_{4}+\sum_{i=1}^3 (\rho_{i}-\rho_{i+1})\, \theta (z-\zeta_{i})\\
     \noalign{\medskip}
     \Sigma(x,z)&=\sum_{i=1}^3\ \sigma_i \,\delta (z-\zeta_{i}) 
     \, ,
     \end{split}
\end{equation}
where, hereafter, 
\begin{equation}\label{sigmadef}
\sigma_i\equiv \rho_{i+1}\ou_{i+1}-\rho_{i}\ou_{i}\, 
\end{equation}
is the horizontal averaged momentum shear.
We remark that field configurations of the form (\ref{I-sigma-rho}) can be regarded as defining a submanifold, which will be denoted 
by~$\MI$,  of Benjamin's Poisson manifold $\MM$ described in Section \ref{Sect-1}.

The $x$ and $z$-derivative of the Benjamin's variables given by equations (\ref{I-sigma-rho}) are 
generalized functions supported at the interfaces  $\{z=\zeta_1\}\cup \{z=\zeta_2\}\cup \{z=\zeta_{3}\}$, and are
computed as
\begin{equation}\label{rhovar}\begin{split}
    \rho_{x}=&-\Sigma_{i=1}^{3} (\rho_{i}-\rho_{i+1})\delta (z-\zeta_{i})\zeta_{ix} \\
    \rho_{z}=&\Sigma_{i=1}^{3}(\rho_{i}-\rho_{i+1})\delta (z-\zeta_{i})\, ,
\end{split}
\end{equation}
and
\begin{equation}\label{dsigma}\begin{split}\medskip
    \Sigma_{x}=& -\Sigma_{i=1}^{3}\sigma_{i}\zeta_{ix}\delta '(z-\zeta_{i})+\Sigma_{i=1}^{3}\sigma_{ix}\delta (z-\zeta_{i})  \\
    \Sigma_{z}=& \Sigma_{i=1}^{3}\sigma_{i}\delta '(z-\zeta_{i})  \, .
\end{split}\end{equation}    
%
To invert the map (\ref{I-sigma-rho}) we choose to integrate along the vertical direction $z$. 
To this end, we define  the two intermediate isopycnals 
%
%
\begin{equation}
\oet_{12}=\dfrac{\zeta_{1}+\zeta_{2}}{2}, \qquad
\oet_{23}=\dfrac{\zeta_{2}+\zeta_{3}}{2}\, .
\end{equation}
\begin{figure}[ht]\label{Figure2}
\centering
\includegraphics[width=.7 \textwidth]{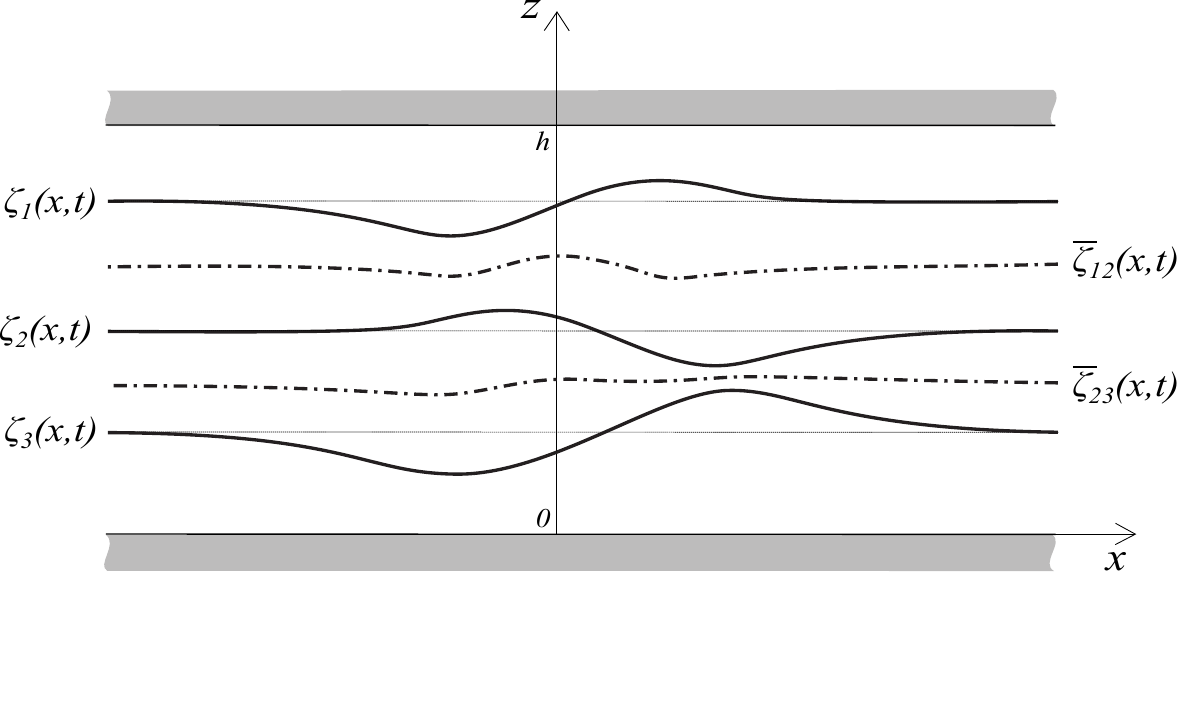}
\caption{Choice of the isopycnals:  $\zeta_i$ are the surface heights and $\oet_{12}, \oet_{23}$ the intermediate isopycnals.}
\label{F2}
\end{figure}
Remarking that $\oet_{12}$ lies in the $\rho_2$-layer and $\oet_{23}$ in the $\rho_3$-layer (see Figure~\ref{F2}), 
by means of this choice
we can introduce on $\MI$ the ``projection"  $\pi$ by
\begin{equation}\label{xitaugen}
\begin{split}
&\pi\left(\rho(x,z), \Sigma(x,z) \right)\equiv \left(\xi_{1},\xi_2,\xi_{3},\tau_1,\tau_2, \tau_3 \right)\\
=&\left( \begin{array}{l}  \int_{0}^{h} (\rho (x,z)-\rho_{4} )\, \D z, \int_{0}^{\oet_{12}} (\rho (x,z) -\rho_{4} )\, \D z, \int_{0}^{\oet_{23}} (\rho (x,z)-\rho_{4} )\, \D z, \\
  \int_{0}^{h} \Sigma (x,z) \D z,\int_{0}^{\oet_{12}} \Sigma (x,z) \D z,\int_{0}^{\oet_{23}} \Sigma (x,z) \D z  
   \end{array} \right)
 \end{split}
   \end{equation}
which maps Benjamin's manifold of $2D$ fluid configurations 
to the space of effective $1D$ fields
$\mathcal{S}$, 
parameterized by the six quantities $(\zeta_k, \sigma_k)$.  A straightforward computation shows that the relations
\begin{equation}\label{xi-sigma}
\begin{split}
\xi_1 &= (h-\zeta_1)(\rho_1 -\rho_2)+(h-\zeta_2) (\rho_2 -\rho_3)+(h-\zeta_3)(\rho_3 -\rho_4) \\
\xi_2 &= \dfrac{\rho_2 -\rho_3}{2}(\zeta_1 -\zeta_2)+\dfrac{\rho_3 -\rho_4}{2} (\zeta_1 +\zeta_2 -2\zeta_3) \\
\xi_{3}&= \dfrac{1}{2}(\rho_3 -\rho_4 )(\zeta_2 -\zeta_3 ) \\
\tau_{1} &= \sigma_{1}+\sigma_{2} + \sigma_{3}, \quad
\tau_{2} = \sigma_{1} +\sigma_{2},\quad
\tau_{3} = \sigma_{3}
\end{split}
\end{equation}
hold.
%



To obtain a Hamiltonian structure on the manifold $\mathcal{S}$ by reducing Benjamin's parent structure~(\ref{B-pb}), we have to perform, as per the Hamiltonian reduction scheme of \cite{Ratiu},  the following steps:
\begin{enumerate}
\item \label{enum} Starting from a 1-form on the 
manifold $\mathcal{S}$, represented by the 6-tuple $$(\alpha_S^{1}, \alpha_S^{2},\alpha_S^{3},\alpha_S^{4},\alpha_S^{5},\alpha_S^{6})\,,$$
we construct its lift  to 
$\mathcal{I}$, that is, a 1-form $\boldsymbol{\beta}_M=(\beta_{\rho},\beta_{\Sigma})$   satisfying the relation
\begin{equation}\label{eqlift}
\int_{-\infty}^{+\infty}\int_0^h( \beta_{\rho} \dot{\rho}+\beta_{\Sigma}\dot{\Sigma})\, \D x\, \D z
= \int_{-\infty}^{+\infty}
\sum_{k=1}^6\alpha_{S}^{k} \cdot \left(\pi_{*} (\dot{\rho},\dot{\Sigma})\right)^k \, \D x\>\>,
\end{equation}
where $\pi_*$ is the tangent map to (\ref{xitaugen}) and $(\dot{\rho},\dot\Sigma)$ are generic infinitesimal variations of $(\rho,\Sigma)$ in the tangent space to $\mathcal{I}$. 
\item
We apply Benjamin's operator~(\ref{B-pb}) to the lifted one form $\boldsymbol{\beta}_M$ 
to get the vector field
\begin{equation}\label{Yfields} 
\left(\begin{array}{c} \dot{\rho}\\
\dot{\Sigma}\end{array}\right)=\left(\begin{array}{c} Y_{M}^{(1)}\\
Y_{M}^{(2)}\end{array} \right)= J_B \cdot \left (\begin{array}{c} 
\beta_{\rho}\\\beta_{\Sigma}
\end{array}\right) 
\, . 
\end{equation}
\item 
 We project the vector $(Y_{M}^{(1)}, Y_{M}^{(2)})$ under
 $\pi_*$ and obtain a vector field  on $\mathcal{S}$. 
 The latter depends linearly on $\{\alpha_S^{(i)}\}_{i=1,\ldots,6}$, and defines
 the reduced Poisson operator $
 {P}$ on $\mathcal{S}$. 
\end{enumerate}
As in the three layer case of \cite{Paper} this construction essentially works as in the two-layer case considered in \cite{CFO17}, provided one  subtle point is taken into account.
Thanks to the relations (\ref{I-sigma-rho}) and the definition of $\pi$, we have that, 
for  tangent vectors $(\dot{\rho},\dot{\Sigma})$,
\begin{equation}\label{pistar1}
\pi_{*}\left(
\begin{array}{c} 
\dot{\rho}\\ \dot{\Sigma}
\end{array}
\right)
=
\left(
\begin{array}{l}\medskip
\int_0^h \dot{\rho}\, \D z\\ \medskip
\int_0^{\oet_{23}} \dot{\rho} \, \D z+\dot{\oet}_{23}\,(\rho(x, \oet_{23})-\rho_4)\\ \medskip
\int_0^{\oet_{12}} \dot{\rho} \, \D z+\dot{\oet}_{12}\,(\rho(x, \oet_{12})-\rho_4)\\ \medskip
\int_0^h \dot{\Sigma}\, \D z\\ \medskip
\int_0^{\oet_{23}} \dot{\Sigma} \, \D z+\dot{\oet}_{23}\, \Sigma(x, \oet_{23})\\ \medskip
\int_0^{\oet_{12}} \dot{\Sigma} \, \D z+\dot{\oet}_{12}\, \Sigma(x, \oet_{12})
\end{array} 
\right)\,.
\end{equation}
Note that in this formula we have an explicit dependence on the variations $\dot{\oet}_{12}$ and $\dot{\oet}_{23}$. To express 
these quantities in terms of $\dot{\rho}$, which is needed to perform the abovementioned steps of the Poisson reduction, 
we can use the analogue of relations~(\ref{rhovar}), that is
\begin{equation}
\label{rhodot}
\dot{\rho}=\Sigma_{i=1}^{3}(\rho_{i+1}-\rho_{i})\dot{\zeta}_{i}\delta(z-\zeta_{i})\, .
\end{equation}
Integrating this with respect to $z$ on the relevant  intervals $[0,h]$, $[0, \oet_{12}]$ and  $[0,\oet_{23}]$ yields
\begin{equation}
\label{A1}\begin{split}
&\int_0^h \dot{\rho}\, \D z=(\rho_4-\rho_3)\dot{\zeta}_3+(\rho_3-\rho_2)\dot{\zeta}_2+(\rho_2-\rho_1)\dot{\zeta}_1\, ,\\
&\int_0^{\oet_{12}} \dot{\rho}\, \D z=(\rho_4-\rho_3)\dot{\zeta}_3+(\rho_3-\rho_2)\dot{\zeta}_2\, , \\ &  \int_0^{\oet_{23}} \dot{\rho}\, \D z=(\rho_4-\rho_3)\dot{\zeta}_3 \, .\end{split}
\end{equation}
Solving the linear system  (\ref{A1}) with respect to the $\dot{\zeta}_k$'s, we can obtain $\dot{\oet}_{12}$ and $\dot{\oet}_{23}$ in terms of integrals of $\dot{\rho}$ along $z$, and thus trade equation~(\ref{pistar1}) for 
\begin{equation}\label{pistar}
\pi_{*}\left(
\begin{array}{c} 
\dot{\rho}\\ \dot{\Sigma}
\end{array}
\right)
=
\left(
\begin{array}{l}\medskip
\int_0^h \dot{\rho}\, \D z\\ \medskip
c_1\int_0^h \dot{\rho}\, \D z + \left( 1+c_3 -c_2\right)\int_{0}^{\oet_{12}} \dot{\rho} \D z -c_3\int_0^{\oet_{23}}\dot{\rho} \D z  \\ \medskip
c_2\int_0^{\oet_{12}}\dot{\rho} \D z +\left( \dfrac{1}{2}-c_2 \right) \int_0^{\oet_{23}}\dot{\rho} \D z \\ \medskip
\int_0^h \dot{\Sigma}\, \D z\\ \medskip
\int_0^{\oet_{12}} \dot{\Sigma} \, \D z \\ \medskip
\int_0^{\oet_{23}} \dot{\Sigma} \, \D z 
\end{array} 
\right) \,,
\end{equation}
where, for the sake of compactness, we use the notation
\begin{equation}\label{cdef}
c_{1}=\dfrac{1}{2}\dfrac{\rho_{2}-\rho_4}{\rho_2 -\rho_1}, \qquad c_2=\dfrac{1}{2}\dfrac{\rho_3 -\rho_4}{\rho_3 -\rho_2}, \qquad c_3 =\dfrac{1}{2}\dfrac{\rho_2 -\rho_4}{\rho_3 -\rho_2}\, .
\end{equation}

We now have at our disposal all the elements to perform the Poisson reduction process.\\
\textbf{Step 1:} The construction of the lifted 1-form $(\beta_{\rho}, \beta_{\Sigma})$ satisfying 
(\ref{eqlift}), i.e.,
\begin{equation}\label{lift2}\begin{split}
&\int_{-\infty}^{+\infty} \int_0^h (\dot{\rho}\, \beta_{\rho}+\dot{\Sigma}\, \beta_{\Sigma})\, \D x\, \D z =
\int_{-\infty}^{+\infty} \Sigma_{k=1}^{6}\alpha_{S}^{k}\pi_{\star}(\dot{\rho},\dot{\Sigma})^k \D x\,, \end{split}
\end{equation}
yields
\begin{equation}\label{liftedform}
\begin{split}
\beta_{\rho}&=\alpha_{S}^{1}+\left(c_1 +(1+c_{3} -c_1)\theta(\oet_{12}-z) -c_3 \theta (\oet_{23}-z)\right)) \, \alpha_{S}^2+
\\ 
&\quad  \left(c_{2} \theta (\oet_{12}-z)+( \dfrac{1}{2}-c_2 )\theta(\oet_{23}-z) \right)\, \alpha_S^3\\\
\beta_{\Sigma}&=\alpha_{S}^4+\theta (\oet_{12}-z)\, \alpha_S^5 + \theta (\oet_{23}-z)\alpha_{S}^6\, .
\end{split}
\end{equation}
In this equation, Heaviside $\theta$'s appear and enable the computation of integrals from the bottom to the chosen isopycnals
$ \oet_{12}$ and $\oet_{12}$  along the full channel $[0,h]$.\\
\textbf{Step 2:} The computation of the vector fields $(Y_{M}^1, Y_{M}^2)$ from the relation 
\begin{equation}\label{Pcomp}
\left(\begin{array}{c} Y_{M}^{(1)}\\
Y_{M}^{(2)}\end{array} \right)= J_B \cdot \left (\begin{array}{c} 
\beta_{\rho}\\\beta_{\Sigma}
\end{array}\right)=\left(\begin{array}{cc}
       0 & \rho_x \partial_z -\rho_z \partial_x \\ 
       \rho_x \partial_z -\rho_z \partial_x & \Sigma_x \partial_z -\Sigma_z \partial_x
      \end{array}\right) \cdot \left (\begin{array}{c} 
\beta_{\rho}\\\beta_{\Sigma}
\end{array}\right) \end{equation}
 is greatly simplified by the specific dependence of the lifted 1-form $(\beta_\rho ,\beta_\Sigma)$ of (\ref{liftedform}) on $z$ and 
 and on the crucial fact that the inequalities  
 $$
 \zeta_{3}<\dfrac{\zeta_{3}+\zeta_{2}}{2}=\oet_{23}<\zeta_2<\oet_{12}=\dsl{\frac{\zeta_1+\zeta_2}{2}}<\zeta_1
 $$  
 hold in the strict sense, so that the terms $\rho_x \partial_z$ and $\Sigma_x \partial_z$ when acting on $(\beta_\rho ,\beta_\Sigma) $ generate products of Dirac $\delta$'s supported at different points, which vanish {\it qua} generalized functions. Moreover,  
 \begin{equation}\label{sigmazbetax}\begin{split}
\Sigma_z\cdot\partial_{x} (\beta_{\Sigma})&=(\Sigma_{i=1}^{3}\sigma_{i}\delta '(z-\zeta_{i})  )\left(\alpha_{S}^4+\theta (\oet_{12}-z)\, \alpha_S^5 + \theta (\oet_{23}-z)\alpha_{S}^6\right)_x\\
&=(\Sigma_{i=1}^{3}\sigma_{i}\delta'(z-\zeta_{i}) )\left(\alpha_{S,x}^4+\theta (\oet_{12}-z)\, \alpha_{S,x}^5 + \theta (\oet_{23}-z)\alpha_{S, x}^6\right)\\ 
&\quad +(\Sigma_{i=1}^{3}\sigma_{i}\delta'(z-\zeta_{i}) )\left(\delta(\oet_{12}-z)\oet_{12,x}\alpha_{S}^5 + \delta (\oet_{23}-z)\oet_{23,x}\alpha_{S}^6\right)\\
&=(\Sigma_{i=1}^{3}\sigma_{i}\delta'(z-\zeta_{i}) )\left(\alpha_{S,x}^4+\theta (\oet_{12}-z)\, \alpha_{S,x}^5 + \theta (\oet_{23}-z)\alpha_{S, x}^6\right)\, ,\end{split}
 \end{equation}
 still due to the above observation  about the supports of the Dirac $\delta$'s. Denoting by $\Delta^{(2)}$ this term, 
 we can write~(\ref{Pcomp}) as  
 \begin{equation}\label{Pcomps}
Y_{M}^{(1)}=-\rho_z(\beta_{\Sigma})_x,\quad Y_{M}^{(2)}=-\rho_z(\beta_{\rho})_x-\Delta^{(2)}\, .
 \end{equation}
 We obtain
\begin{equation}
\begin{split}
&Y_{M}^{(1) }=\left(\sum_{k=1}^3(\rho_k -\rho_{k+1})\delta (z-\zeta_k)\right)\alpha_{S,x}^4
\\&+\left(\sum_{k=2}^3(\rho_k -\rho_{k+1})\delta (z-\zeta_k)\right)\alpha_{S,x}^5+
 (\rho_3 -\rho_4)\delta (z-\zeta_3)\alpha_{S,x}^6 \,,
\end{split}
\end{equation}
as well as the more complicated formula for  $Y_{M}^{(2)}$,
\begin{equation}\label{Y2M}
Y_{M}^{(2)} =(\sum_{i=1}^3(\rho_i -\rho_{i+1} )\delta (z-\zeta_i))\, (\alpha_{S,x}^1+K_2 \alpha_{S,x}^2+K_3\alpha_{S,x}^3)-\Delta^{(2)}\, , 
\end{equation}
where
\begin{equation}\label{Kdef}
\begin{split}
K_2&= c_{1} +(1+c_3 -c_{1})\theta (\oet_{12}-z)-c_3 \theta(\oet_{23}-z)\\
&K_3=c_2 \theta (\oet_{12}-z)+\left( \dfrac{1}{2}-c_2 \right) \theta (\oet_{23}-z) \,.
\end{split}
\end{equation}
\textbf{Step 3:} The computation of the push-forward under the map $\pi_*$ of the vector field $(Y_{M}^{(1)}, Y_{M}^{(2)})$, to obtain the six-component vector field $(\dot{\xi_k}, \dot{\tau}_k)$ on $\mathcal{S}$ is a direct but tedious task. 
Thanks to the explicit expressions (\ref{cdef}) and (\ref{Kdef}), substituting in (\ref{pistar}) and noticing that, due to the presence of the $z$-derivatives of the Dirac $\delta$, $\Delta^{(2)}$ is in the kernel of $\pi_*$,  yields
\begin{equation}\label{pushfwd}
\begin{split}
\dot{\xi}_1&=\alpha_{S,x}^4 (\rho_1 -\rho_4)+\alpha_{S,x}^5 (\rho_2 -\rho_4)+\alpha_{S,x}^6 (\rho_3 -\rho_4)\\
\dot{\xi}_2&=\dfrac{1}{2}(\rho_{2}-\rho_{4})\alpha_{S,x}^{5}+ (\rho_3 -\rho_4)\alpha_{S,x}^6\\
\dot{\xi}_3&=\dfrac{1}{2}(\rho_{3}-\rho_{4})\alpha_{S,x}^6\\
\dot{\sigma}_1&=(\rho_1 -\rho_4)\alpha_{S,x}^{1}\\
\dot{\sigma}_2&=(\rho_2 -\rho_4)\alpha_{S,x}^1 +\dfrac{1}{2}(\rho_2 -\rho_4)\alpha_{S,x}^2 \\
\dot{\sigma}_3&=(\rho_3 -\rho_4)\alpha_{S,x}^1 +(\rho_3 -\rho_4)\alpha_{S,x}^2+\dfrac{1}{2}(\rho_3 -\rho_4)\alpha_{S,x}^3 \,.
\end{split}
\end{equation}
Thus, the Poisson tensor $P
$ on the manifold $\mathcal{S}$ in the coordinates $(\xi_1 ,\xi_2 ,\xi_3 , \tau_1 ,\tau_2 ,\tau_3)$   becomes
\[P
=\left( {\begin{array}{*{20}{c}}
0&A\\
{{A^T}}&0
\end{array}} \right){\partial _x}\,,\text{
where }
A = \left( {\begin{array}{*{20}{c}}\medskip
{\rho_1 -\rho_4}&{\rho_2 -\rho_4}&{\rho_3 -\rho_4}\\
{0}&{\dfrac{\rho_2 -\rho_4}{2}}&{\rho_3 -\rho_4}\\
{0}&{0}&{\dfrac{\rho_3 -\rho_4}{2}}
\end{array}} \right) \, . \]
Recalling  relations (\ref{xi-sigma}), 
a straightforward computation  shows that in the coordinates 
$(\zeta_1 ,\zeta_2 ,\zeta_3 ,\sigma_1 ,\sigma_2 ,\sigma_3)$ 
the reduced Poisson operator acquires the particularly simple form 
\begin{equation}
\label{PoissonAveraged}
P
= \left( {\begin{array}{*{20}{c}}
0&0&0&{ - 1}&0&0\\
0&0&0&0&{ - 1}&0\\
0&0&0&0&0&{ - 1}\\
{ - 1}&0&0&0&0&0\\
0&{ - 1}&0&0&0&0\\
0&0&{ - 1}&0&0&0
\end{array}} \right){\partial _x}\,.
\end{equation}
\begin{remark}\label{remarks1} 
According to the terminology favored by the Russian school, for Hamiltonian quasi-linear systems of PDEs  
the coordinates $(\xi_l, \tau_l)$ and, {\em a fortiori}, the coordinates $(\zeta_l, \sigma_l)$, are ``flat" coordinates for the system. In view of the particularly simple form of the Poisson tensor (\ref{PoissonAveraged}), the latter set could be called a system of flat {\em Darboux} coordinates. 
\end{remark}

\begin{remark}\label{remarks2}
In \cite{Paper} we conjectured that in  the $n$-layered case, with a stratification given by densities $\rho_1<\rho_2<\cdots<\rho_n$ and interfaces $\zeta_1>\zeta_2>\cdots>\zeta_{n-1}$, a procedure yielding a natural Hamiltonian formulation for the averaged problem was to  consider intervals
\begin{equation}\label{N-int}
I_1=[0,h],\, I_2=\left[0, \frac{\zeta_1+\zeta_2}{2}\right], I_3=\left[0, \frac{\zeta_2+\zeta_3}{2}\right],\,\dots\,,\, I_n=\left[0, \frac{\zeta_{n-2}+\zeta_{n-1}}{2}\right]\,.
\end{equation}
We explicitly proved it here  for $n=4$, together with the conjecture that the quantities
\begin{equation}\label{conj-N}
(\zeta_1,\zeta_2,\zeta_{3}, \sigma_1,\sigma_2,\sigma_{3}), 
\end{equation}
 where $\sigma_k=\rho_{k+1}\ou_{k+1}-\rho_{k}\ou_{k}$, are  flat  Darboux coordinates for the reduced Poisson structure. 
 \end{remark}
\section{The reduced Hamiltonian under the Boussinesq approximation}
\label{redham}
The  energy  of the $2D$ fluid in the channel is just the sum of the kinetic and potential energy, 
\begin{equation}\label{E2D}
H=\int_{-\infty}^{+\infty} \int_0^h \frac\rho2\ 
\left(u^2+w^2\right) \, \D x\, \D z+\int_{-\infty}^{+\infty} \int_0^h  g(\rho-\rho_0) 
z\, \D x\, \D z\,, 
\end{equation}
where $\rho_0$ is the reference density fixed by the far field constant values of the layers' thicknesses. In our case we have 
$\rho_0=\sum_{i=1}^4 \rho_i\eta_i^{(\infty)}\, ,$ where  $\eta_i^{(\infty)}$ are the asymptotic values of the $\eta_i$'s as $|x|\to \infty$.

The potential energy  is thus readily reduced, using the first of 
\rref{rhouw}, to 
\begin{equation}\label{Ured}
U= \int_{-\infty}^{+\infty}  \frac12 \left(\,g \left( \rho_{{2}}-\rho_{{1}} \right) {\zeta_{{1}}}^{2}+
g \left( \rho_{{3}}-\rho_{{2}} \right) {\zeta_{{2}}}^{2}+g \left(\rho_4 -\rho_3 \right)\zeta_{3}^{2}\right)\, \D x +U_\Delta \,, 
\end{equation}
where $U_\Delta $ contains  constant and  linear in the $\zeta_k$'s terms,  which ensure the convergence of the integral, but that do not affect the equations of  motion in view of the form (\ref{PoissonAveraged}) of the Poisson tensor.

%
To obtain the reduced kinetic energy density, we use the fact that at order~$O(\epsilon^2)$ we can disregard the vertical velocity $w$, and trade the horizontal velocities with their layer-averaged means. Thus the $x$-density is computed as
\begin{equation}\label{Tred}
\begin{split}
\mathcal{T}&=\frac12\left(\int_0^{\zeta_3} \rho_4 \ou_4^2 \, \D z+\int_{\zeta_3}^{\zeta_2}\rho_3 \ou_3^2\, \D z+\int_{\zeta_2}^{\zeta_1}  \rho_2 \ou_2^2\, \D z +\int_{\zeta_1}^{h}  \rho_1 \ou_1^2\, \D z\right)\\
&=\frac12\left( \rho_4{\zeta_3}\ou_4^2 
+ \rho_3({\zeta_2-\zeta_3})\ou_3^2+ \rho_2{(\zeta_1 -\zeta_2)}\ou_2^2+ \rho_1{(h-\zeta_1)}\ou_1^2\right)\, .
\end{split}
\end{equation}

The so-called Boussinesq approximation consists of the double scaling limit 
\begin{equation} \label{Baxx}
\rho_i\to\bar{\rho},\,\, i=1,\ldots, 4, \quad g\to\infty \text{ with } g(\rho_{j+1}-\rho_j)\, \text{ finite}, \,\, j=1,2,3 .
\end{equation}
where 
$$
\bar{\rho}=\frac14 \,\,\dsl{\sum_{i=1}^4\rho_i}
$$ 
denotes the average density. 
This approximation then consists of neglecting density differences in the inertia terms of stratified Euler fluids, while retaining these differences in the buoyancy terms, owing to the relative magnitude of gravity forces with respect to those from inertia. This results in the Boussinesq  energy density 
\begin{equation}\label{TredB}
\begin{split}
\mathcal{E} &=\frac{\bar{\rho}}2\left(\zeta_3\ou_4^2 
+ (\zeta_2-\zeta_3)\ou_3^2+ (\zeta_1 -\zeta_2)\ou_2^2+ (h-\zeta_1)\ou_1^2\right)\\ 
& \quad +\frac12 \left(\,g \left( \rho_{{2}}-\rho_{{1}} \right) \zeta_1^{2} +
g \left( \rho_{{3}}-\rho_{{2}} \right) \zeta_{{2}}^{2}+g \left(\rho_4 -\rho_3 \right)\zeta_{3}^{2}\right)\, .
\end{split}
\end{equation}
To express this energy in terms of the Hamiltonian variables $(\zeta_i, \sigma_i)$, $i=1,2,3$, 
we use the dynamical constraint
\begin{equation}\label{dynconstr}
(h-\zeta_1)\ou_1+(\zeta_1-\zeta_2)\ou_2+(\zeta_2 -\zeta_3)\ou_3+\zeta_3 \ou_4=0\, ,
\end{equation}
as well as the definitions (\ref{sigmadef}) that, in the Boussinesq approximation, are turned into
\begin{equation}\label{Bdefsigma}
\sigma_k=\bar{\rho}(\ou_{k+1}-\ou_k)\,.
\end{equation}
We get
\begin{equation}\label{szetatou}
\begin{split}
\ou_1&= -\frac{\zeta_1 \sigma_1+\zeta_2 \sigma_2+\zeta_3 \sigma_3}{h \bar{\rho} },\\
\ou_2&= -\frac{\zeta_1 \sigma_1+\zeta_2 \sigma_2+\zeta_3 \sigma_3-h \sigma_1}{h \bar{\rho} },\\
\ou_3&= -\frac{\zeta_1 \sigma_1+\zeta_2 \sigma_2+\zeta_3 \sigma_3-h \sigma_1-h \sigma_2}{h \bar{\rho} },\\
\ou_4&= -\frac{\zeta_1 \sigma_1+\zeta_2 \sigma_2+\zeta_3 \sigma_3-h \sigma_1-h \sigma_2-h \sigma_3}{h \bar{\rho} }\, .
\end{split}
\end{equation}
Hence, from (\ref{TredB}), the Hamiltonian functional 
%
acquires its final form in the Boussinesq approximation as 
\begin{equation}
\label{FullH}
\begin{split}
H_B=&
\int_{\RR} 
\left( \frac1{2\, h\bar{\rho}}  \, 
\left( \sigma_{1}^{2} \zeta_{1} 
\left(h -\zeta_{1}\right) +\sigma_{2}^{2} 
\left(h -\zeta_{2}\right) \zeta_{2} +\sigma_{3}^{2} 
\left(h -\zeta_{3}\right) \zeta_{3}+
\right.
\right. 
\\ 
&\left.\left.
2 \sigma_{1} \sigma_{2} \zeta_{2} 
\left(h -\zeta_{1}\right)
+2 \sigma_{1} \sigma_{3} \zeta_{3} \left(h -\zeta_{1}\right)
+2 \sigma_{2} \sigma_{3}\zeta_{3}  \left(h -\zeta_{2}\right) 
\right) +\right.
\\  & \left. \frac{g}{2} \left( (\rho_2 - \rho_1) \zeta_{1}^2 +  (\rho_3 -\rho_2) \zeta_{2}^{2}+(\rho_4 -\rho_3)\zeta_{3}^{2}\right)
\right)\, \D x\, .
\end{split}
%
%
%
 \end{equation}
Thanks to the simple form of the Poisson tensor  (\ref{PoissonAveraged}), the ensuing equations of motion can be written as the conservation laws
 \begin{equation}\label{equofmot}
\begin{array}{l}\medskip \dsl{
{\zeta_1}_t+\left(\frac{\sigma_{1} \zeta_{1} \left(h -\zeta_{1}\right)}{h \rho}+\frac{\sigma_{3} \zeta_{3} \left(h -\zeta_{1}\right)}{h \rho}+\frac{\sigma_{2} \zeta_{2} \left(h -\zeta_{1}\right)}{h \rho}\right)_x=0}\\\medskip
\dsl{{\zeta_2}_t+\left(\frac{\sigma_{2} \left(h -\zeta_{2}\right) \zeta_{2}}{h \rho}+\frac{\sigma_{3} \left(h -\zeta_{2}\right) \zeta_{3}}{h \rho}+\frac{\sigma_{1} \zeta_{2} \left(h -\zeta_{1}\right)}{h \rho}\right)_x=0}
\\ \medskip
\dsl{{\zeta_3}_t+\left(\frac{\sigma_{3} \left(h -\zeta_{3}\right) \zeta_{3}}{h \rho}+\frac{\sigma_{2} \left(h -\zeta_{2}\right) \zeta_{3}}{h \rho}+\frac{\sigma_{1} \zeta_{3} \left(h -\zeta_{1}\right)}{h \rho}\right)_x=0}\\ \medskip
\dsl{{\sigma_1}_t+\left(\frac{\left(h -2 \zeta_{1}\right) \sigma_{1}^{2}}{2 h \rho}-\frac{\sigma_{1} \sigma_{2} \zeta_{2}}{h \rho}-\frac{\sigma_{1} \sigma_{3} \zeta_{3}}{h \rho}+g \left(\rho_{2}-\rho_{1}\right) \zeta_{1}\right)_x=0}\\\medskip
\dsl{{\sigma_2}_t+\left(\frac{\left(h-2 \zeta_{2} \right) \sigma_{2}^{2}}{2 h \rho}-\frac{\sigma_{2} \sigma_{3} \zeta_{3}}{h \rho}+\frac{\sigma_{1} \sigma_{2} \left(h -\zeta_{1}\right)}{h \rho}+g \left(\rho_{3}-\rho_{2}\right) \zeta_{2}\right)_x=0}\\
\dsl{{\sigma_3}_t+\left(\frac{\left(h-2 \zeta_{3} \right) \sigma_{3}^{2}}{2 h \rho}+\frac{\sigma_{2} \sigma_{3} \left(h -\zeta_{2}\right)}{h \rho}+\frac{\sigma_{1} \sigma_{3} \left(h -\zeta_{1}\right)}{h \rho}+g \left(\rho_{4}-\rho_{3}\right) \zeta_{3}\right)_x=0} \,.
 \end{array}
\end{equation}
The Hamiltonian formalism easily shows the existence of the eight conserved quantities
\begin{equation}\label{consqua}
\begin{split} 
&Z_j=\int_{-\infty}^{+\infty}  \zeta_j\, \D x,\qquad S_j=\int_{-\infty}^{+\infty}  \sigma_j\, \D x, \quad j=1,2,3\, \, , \\
&K=\int_{-\infty}^{+\infty}  \sum_{k=1}^3 \zeta_k\sigma_k\, \D x\quad \text{ and }  H_B \text{ given by (\ref{FullH}).}
\end{split}
\end{equation}
\begin{remark}
The first six quantities are Casimir functionals for the Darboux Poisson tensor (\ref{PoissonAveraged}), while the seventh one, $K$, is the generator of $x$-translations.
Note that, formulas (\ref{szetatou}) imply  that  the total linear momenta of the individual layers are conserved quantities. 
This is consistent with the fact that the dispersionless limit of the $N$-layer equations are conservation laws for the averaged momenta, and no pressure imbalances can arise in the Boussinesq approximation \cite{Paradox}.
\end{remark}
\begin{remark}\label{remark3} 
The steps leading to the computation of the effective Hamiltonian  (\ref{FullH}) can be performed also by dropping the assumptions (\ref{Baxx}) of the Boussinesq approximation. In this case, the kinetic energy acquires a non trivial rational dependence on the density differences $\rho_i-\rho_{i+1}$, and the equations of motion become much more complicated (as already seen  in the  $2$ and $3$-layer cases). However, they are still Hamiltonian equations of motion that preserve, together with their Hamiltonian, 
the quantities $Z_j,\, S_j,\, j=1,2,3$ and the generator of $x$-translations $K$ of eq. (\ref{consqua}). 
Note that, as shown in \cite{Paradox} and further discussed in \cite{Paper}, once beyond the Boussinesq approximation pressure imbalances can appear. Hence the
individual layer momenta are no longer conserved quantities and  $K$ does not even coincide with the total horizontal momentum.
\end{remark}


\section{Symmetric solutions} \label{SySo}
Symmetric solutions of the three-layer configurations were ingeniously found in \cite{VirMil19} by a direct inspection of the equations of motion (written in velocity -- thickness coordinates). They exist provided a certain relation is enforced on the density differences of the individual layers, and were interpreted in \cite{Paper} as the fixed point of a suitable canonical involution of the phase space of the 3-layer model. 

Here we shall adopt the latter point of view, and identify an involution of the phase space of the $4$-layer model above that leads to the existence of a family of symmetric solutions.
First, we focus on the kinetic energy part of the Boussinesq model~(\ref{FullH}), 
\begin{equation}
\label{TB}
\begin{split}
\mathcal{T}_B=
&\frac1{2\, h\bar{\rho}}\, 
\Big( \sigma_{1}^{2} \zeta_{1} 
\left(h -\zeta_{1}\right)
+\sigma_{2}^{2} \left(h -\zeta_{2}\right) \zeta_{2} 
+\sigma_{3}^{2} \left(h -\zeta_{3}\right) \zeta_{3}+  \\ 
&
2 \sigma_{1} \sigma_{2} \zeta_{2} \left(h -\zeta_{1}\right)
+2 \sigma_{1} \sigma_{3} \zeta_{3} \left(h -\zeta_{1}\right)
+2 \sigma_{2} \sigma_{3} \left(h -\zeta_{2}\right) \zeta_{3}  \Big) .
\end{split}
\end{equation}
This expression is clearly invariant under the involutive map
\begin{equation}
\label{invo}
\zeta_1\to h-\zeta_3,\quad \zeta_2\to h-\zeta_2,\quad \zeta_3\to h-\zeta_1,\quad \sigma_1\to -\sigma_3,\quad \sigma_2 \to -\sigma_2, 
\quad\sigma_3\to-\sigma_1\, .
\end{equation}
If we assume that the densities $\rho_k$  fulfill the relations
\begin{equation}
\label{relrho}
\rho_4-\rho_3=\rho_2-\rho_1\equiv\drho\, , 
\end{equation}
the Hamiltonian density  (\ref{FullH}) is invariant as well, up to the addition of linear terms in the $\zeta$'s, that is, up to  constant terms and Casimir densities of the Poisson tensor  $P
$ of (\ref{PoissonAveraged}) which do not affect the equations of motion.
A straightforward computation shows that the Poisson tensor (\ref{PoissonAveraged}) is left invariant by the above involution. 
Hence,  the manifold $\mathcal{F}$ of fixed points of the involution (\ref{invo}) is invariant under the Hamiltonian flow (\ref{equofmot}).

The above statement can be cast in a more geometrical light. 
Suppose that we are given a Poisson manifold $(M,P)$ with Hamilton equations written generically as
\begin{equation}
{z}_t=P\, dH\, , 
\end{equation}
and suppose that $z\to \varphi(z)$ is an involution preserving $P$, i.e.,
\begin{description}\label{propinv}
\item[i) ] $ \varphi\circ \varphi=\mathrm{Id}$\\
\item[ii)]$\varphi_*\, P\, \varphi^* =P$, where $\varphi_*$ is the (Fr\'echet) derivative of $\varphi$, and $\varphi^*$ is its pull-back (from the linear algebra perspective, the adjoint map).

\end{description}
Then 
\begin{equation}\label{theoinv}
{\varphi(z)}_t=\varphi_* {z}_t=\varphi_*P\, dH=\varphi_*P\varphi^*\varphi^*dH=P\varphi^*dH=P d \varphi^* H\,.
\end{equation}
Hence, if $z$ satisfies $\varphi(z)=z$ we have ${\varphi(z)}_t-{z}_t=0$ so that initial data  fixed by the involution $\varphi$ remain on the invariant submanifold during the time evolution.
In our case, the invariant  manifold can be explicitly described as the submanifold of $\mathcal{S}$ characterised by the constraints
\begin{equation}
\label{Fconst}
\zeta_1+\zeta_3-h=0\,, \quad \zeta_2-\frac{h}{2}=0\,,\quad \sigma_1+\sigma_3=0\,, \quad \sigma_2=0
\,, 
\end{equation}
and is parametrized by two of the remaining variables, for instance the two quantities
\begin{equation}
\label{Fvar}
\sigma\equiv\sigma_3\,,\quad  \zeta\equiv\zeta_3.
\end{equation}
The reduced equations of motion on $\mathcal{F}$  in these variables are 
\begin{equation}
\label{eomr}
\left\{\begin{array} {l}\medskip
\dsl{\zeta_t-\frac{2(\zeta^2\sigma)_x}{h\bar{\rho}}+\frac{(\zeta\sigma)_x}{\bar{\rho}}=0}\\
\dsl{ \sigma_t+\frac12\frac{((h-4\zeta)\sigma^2)_x}{h\bar{\rho}} +2g\drho \zeta\zeta_x=0 }
\end{array}
\right. \, , 
\end{equation}
while the restriction of the Hamiltonian (\ref{FullH}) is
\begin{equation}
\label{HF}
H_\mathcal{F}=\int_\RR \left(\frac{\zeta \left(h -2\zeta\right) \sigma^{2}}{h \bar{\rho}}+g \rho_{\Delta} \zeta^{2}\right)\, \D x\,.
\end{equation}
\begin{figure}[ht]
\centering
\includegraphics[width=.7 \textwidth]{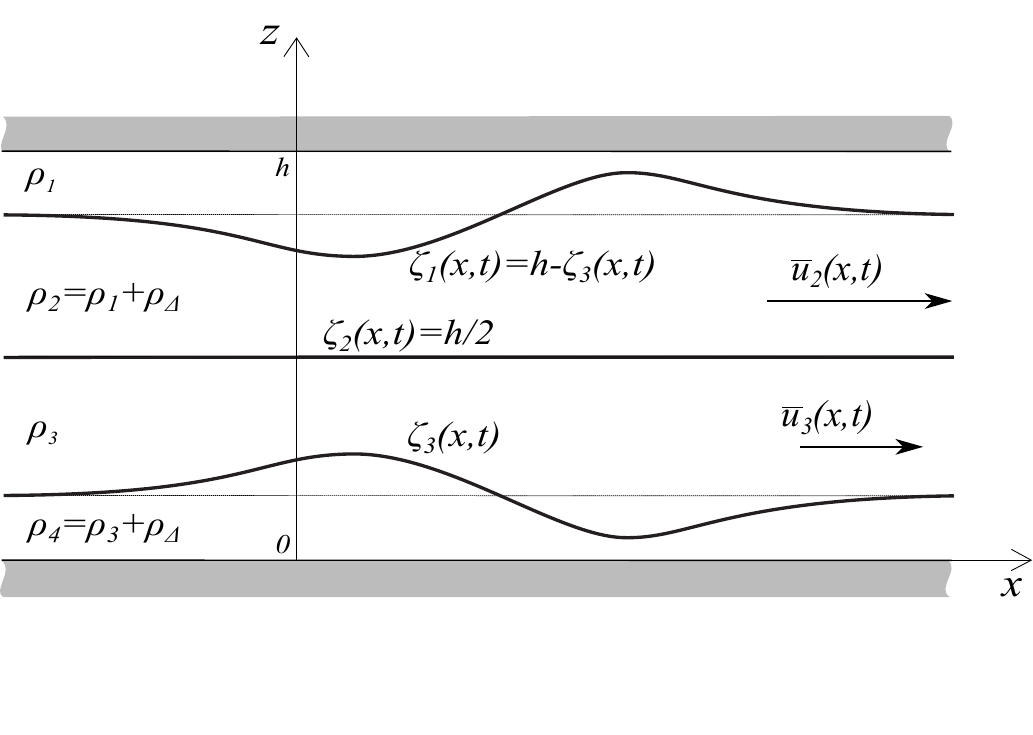}
\caption{Example of a symmetric solution.}
\label{F3}
\end{figure}
One can readily check the equations (\ref{eomr}) are the Hamiltonian equations of motion
\begin{equation}
\label{FHeom}
\left(
\begin{array}{c}\medskip
\zeta_t\\\sigma_t
\end{array}
\right)
=P_\mathcal{F}\left(
\begin{array}{c}\medskip
\delta_\zeta H_\mathcal{F}\\
\delta_\sigma H_\mathcal{F}
\end{array}
\right)
\, 
\text{  with } P_\mathcal{F}=\left(
\begin{array}{cc}\medskip
0&-\dsl{\frac12{\partial_x}}\\
-\dsl{\frac12{\partial_x}}&0
\end{array}
\right)\, .
\end{equation}
The appearance of the factor $1/2$ in the expression of $P_\mathcal{F}$ is readily explained within Dirac's  theory of constrained Hamiltonian systems.
Indeed, if we consider the constraints (\ref{Fconst}), we notice that, renaming the constraint densities as
\begin{equation}
\Phi_1=\zeta_1+\zeta_3-h\,,\quad \Phi_2=\zeta_2-h/2\,,\quad \Phi_3=\sigma_1+\sigma_3\,, \quad \Phi_4=\sigma_2 
\, ,
\end{equation}
the sixtuple $(\zeta=\zeta_3, \sigma=\sigma_3, \Phi_1, \ldots, \Phi_4)$ is clearly a set of coordinates.
The Poisson tensor in these coordinates is given by the block matrix
\begin{equation}\label{Pnew}
P=-\partial_x \left(\begin{array}{cc} A&B^T\\B&C\end{array}\right)\, ,
\end{equation}
with
\begin{equation}
A=\left(\begin{array}{cc}0&1\\1&0\end{array}\right),\quad B=\left(\begin{array}{cc}0&1\\0&0\\1&0\\0&0\end{array}\right),\quad 
C=\left(\begin{array}{cccc}0&0&2&0\\0&0&0&1\\ 2&0&0&0\\0&1&0&0\end{array}\right)\, .
\end{equation}
In this formalism, Dirac's formula \cite{Dirac} for the $2\times 2$ reduced tensor $P^D$ with respect to the pair of coordinates $(\zeta, \sigma)$ on the constrained manifold  is
\begin{equation}\label{PD}
P^D=\left(A-B^T\cdot C^{-1}\cdot B\right)\partial_x \, , 
\end{equation}
by which we recover the tensor $P_\mathcal{F}$ of (\ref{FHeom}).

As a final remark, we notice that the symmetric solutions might appear somewhat trivial, especially  in view of the  constraint that fixes the intermediate height $\zeta_2$ to be at the middle of the channel. However, one should remark that the additional requirement (\ref{relrho}) on the densities reads
$\rho_3=\rho_4 -\rho_2+\rho_1$. Thus in general $\rho_3\neq \rho_2$ and, in the non-Boussinesq case, the constraint 
$\sigma_2=\rho_3\ou_3-\rho_2\ou_2=0$ generates a velocity shear along the flat interface $\zeta_2$.
\section{Conclusions and discussion} \label{CoDi}
Building on our previous paper \cite{Paper}, we have considered the reduction of a natural Hamiltonian structure from the space of $2D$ general stratified configurations for an Euler incompressible, non-homogeneous fluid, to an effective $1D$ model of a four-layered sharply stratified fluid, in the long-wave dispersionless approximation. We have applied a general scheme for reducing Hamiltonian structures and constructed a set of natural Darboux coordinates on the reduced space of sharply stratified configurations.
We then constructed an effective Hamiltonian in the Boussinesq approximation setting, which basically retains density differences only in the buoyancy terms. We finally pointed out the existence of a special family of symmetric solutions to the effective equations of motion which generalize their 3-layer counterpart and can serve to illustrate the expected differences and analogies between the general cases of odd and even number of layers.
Symmetric solutions in the $3$-layer case were introduced in \cite{VirMil19} as a natural setting for the so-called mode 2 internal waves, that is, internal waves with out-of-phase pycnocline displacements. We have geometrically shown that  four layer stratifications can support a similar family of waves, with the notable peculiarity that such ``opposite" disturbances in the pycnocline displacement happen in the first and fourth layer only. While the middle interface $\zeta_2$  is forced to be flat and at the middle of the channel,  velocity shears along this interface are not ruled out. Analytical and numerical properties of these $4$-layers solutions (e.g., stability and well-posedness) are currently under investigation and will be communicated elsewhere.

%
%
%
%
%

\section*{Acknowledgments}
TTVH would like to thank the organizers of the 
NMMP-2022 Workshop held virtually in Tallahassee, June 17--19, 2022,  for  the opportunity to present results related with the topic of this paper, as well as the UNC Mathematics Department and CCIAM
for hosting her visit in the Spring 2022. 
This project has received funding from the European Union's Horizon 2020 research and innovation programme under the Marie Sk{\l}odowska-Curie grant no 778010 {\em IPaDEGAN}. We also gratefully acknowledge the auspices of the GNFM Section of INdAM, under which part of this work was carried out, and the support of the project MMNLP (Mathematical Methods in Non Linear Physics) of 
INFN. RC thanks the support by the National Science Foundation under grants RTG DMS-0943851, CMG ARC-1025523, DMS-1009750, DMS-1517879, DMS-1910824, and by the Office of Naval Research under grants N00014-18-1-2490 and DURIP N00014-12-1-0749. RC \& MP thank the Department of Mathematics and Applications of the University of Milano-Bicocca for its hospitality. 
%
%

\end{document}